\documentclass[epj]{webofc}
\usepackage[varg]{txfonts}   
\usepackage{hyperref}

\renewcommand*{\eqref}[1]{Eq.~(\ref{eq:#1})}

\graphicspath{{./fig/}}

\wocname{VLVNT2008}
\woctitle{VLVNT2008}
\begin{document}
\title{Radio detection in the multi-messenger context}
%
%

\author{
\firstname{D.}~\lastname{Kostunin}\inst{1}
}


\institute{
DESY, 15738 Zeuthen, Germany
}

\abstract{The present work discusses the development of the radio technique for detection of ultra-high energy air-showers induced by cosmic radiation,
and the prospects of its application in the future multi-messenger activities, particularly for detection of ultra-high energy cosmic rays, gamma rays and neutrinos.
It gives an overview of the results achieved by the modern digital radio arrays, as well as discuss present challenges and future prospects.}

\maketitle
\section{Introduction}
Radio detection of ultra-high energy ($>$PeV) particle cascades in media was proposed~\cite{Askaryan:1962hbi, KahnLerche1966} and detected~\cite{jelley} more than half of century ago.
Due to limitations of the data acquisition and data analysis with the technologies available at that time, the radio technique has been disregarded until the first decade of the 21st century.
Development and mass production of fast digital-analog converters, boards and computers enabled the installation of large wide-angle digital radio arrays for radio astronomy as well as for air-shower detection.
In the last years digital radio arrays operating in the MHz frequency band have proven their feasibility, hardware and software for them were developed and successfully applied.
The interest in radio detection of air-showers was rekindled due to the following features of this technique:
\begin{itemize}
\item \emph{Cost-efficiency}.
The cost of a single detection element (antenna) of a radio array are an order of magnitude lower than for particle detectors (scintillators) and optical detectors (PMTs).
At the same time, deployment and maintenance of radio arrays require less human, time and financial resources than for optical arrays or telescopes.
\item \emph{Duty-cycle}.
Since the air is transparent for MHz radio, the detection is almost unaffected by the atmospheric conditions (temperature, density and humidity, see Ref.~\cite{Corstanje:2017djm} for details) and can be performed around-the-clock except during thunderstorms.
\item \emph{Precision for energy and shower maximum}.
In the last years it was proven that the resolution of radio detectors can achieve 10-15\% for the energy and 20-40~g/cm\textsuperscript{2} for the depth of shower maximum~\cite{Buitink:2014eqa,Bezyazeekov:2018yjw} depending on energy and on the configuration of the detector.
These numbers are comparable with the precision achievable using optical methods of air-shower detection.
\item \emph{Sensitivity for inclined events}.
Since the secondary particles as well as Cherenkov and fluorescent light are absorbed during the propagation through the atmosphere, optical and particle setups have difficulties to detect very inclined air showers (with inclination $\theta>60^\circ$) with full efficiency.
Contrary to it, radio waves can propagate tens of kilometers in the atmosphere and be seen by an antenna array from a very far distance.
Although, the power of the emission falls with distance squared, the air-shower footprint increases as $1/\cos\theta$, which allows one to detect these air-showers with very sparse arrays.
\end{itemize}
The combination of these features makes the detection of ultra-high energy messengers ($>$\,EeV) the perfect science case for the radio technique.
For the time being most of the digital radio arrays serve as extensions for the existing cosmic-ray setups, and only few operate in stand-alone mode.
The main obstacle for large-scale stand-alone array is the high radio background.
To achieve high efficiency under this background one needs to select the optimal frequency band and develop sophisticated \emph{self-trigger}.
In the present paper I discuss the actual progress in solving this problem and have an overview of the proposed radio experiments which could contribute to ultra-high energy multimessenger.

\section{Progress in the detection of air-showers with radio}

\begin{figure}[t]
\centering
\includegraphics[height=0.45\textwidth]{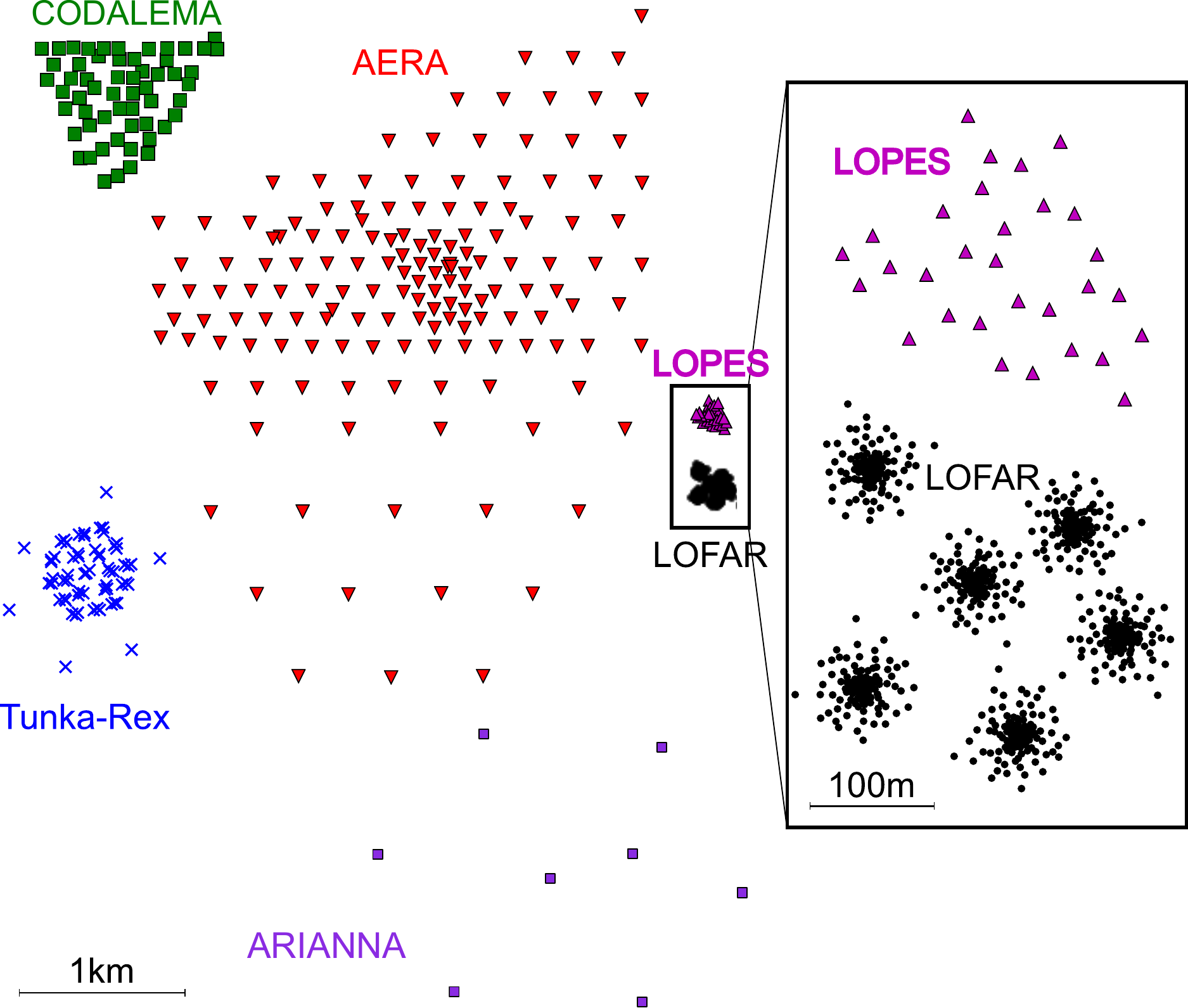}~~~~~~
\includegraphics[height=0.45\textwidth]{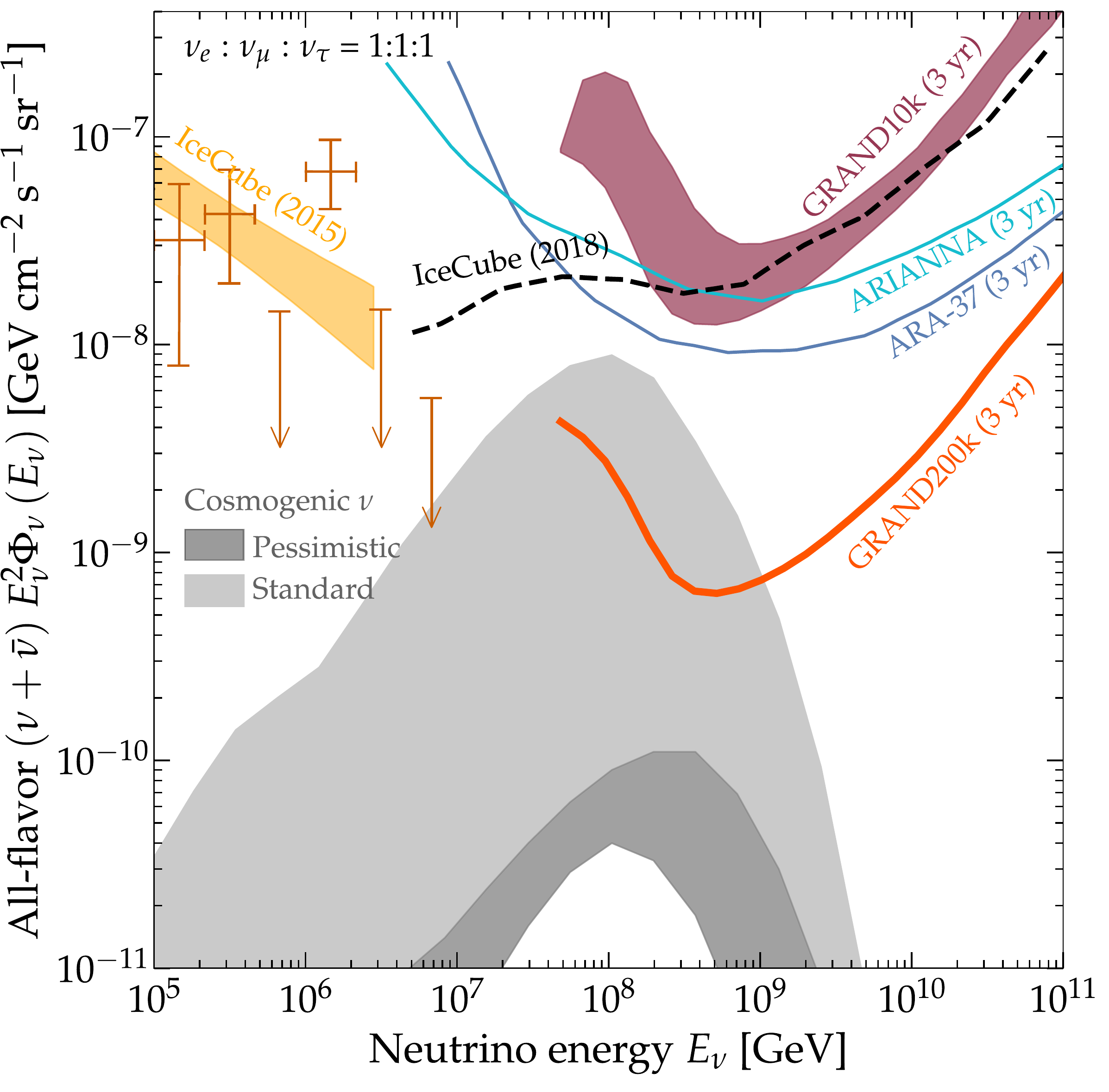}
\caption{\textit{Left:} Comparison of layouts of radio arrays discussed in the present paper (courtesy of A. Zilles).
\textit{Right:} Sensitivity of the present and proposed neutrino detectors compared to the predicted ultra-high energy neutrino fluxes~\cite{AlvesBatista:2018zui}.
IceCube limits and points are taken from~\cite{Aartsen:2015knd, Aartsen:2018vtx}.
The plot is adapted from the GRAND whitepaper~\cite{Alvarez-Muniz:2018bhp}.
}
\label{fig:all_layouts}
\end{figure}

The proof-of-principle of the detection of air showers with digital arrays was provided by LOPES~\cite{FalckeNature2005} and CODALEMA~\cite{Ardouin2005148} in the middle of the 2000s.
They had shown the possibility of the reconstruction of air-shower direction and primary energy with promising resolution.
For a more precise study of the phenomena, it was necessary to have simulation packages describing all known features with sufficient accuracy.
For this purposes CoREAS~\cite{Huege:2013vt}, a CORSIKA~\cite{HeckKnappCapdevielle1998} extension was developed and released in 2012.
From the experimental side a great contribution to the field was done with LOFAR~\cite{LOFAR_general}, a digital radio array designed for astronomy purposes and a cosmic-ray mode operating jointly with co-located particle detectors.
With its dense layout, LOFAR has confirmed good agreement between theory and observations for the 30-80 MHz band~\cite{Buitink:2014eqa}, as well as for higher frequencies~\cite{Nelles:2014dja}.
Unfortunately, such dense arrays cannot be used for the studying air showers produced by particles with energies higher than $10^{17.5}$~eV due to their small area.
To acquire sufficient data in the EeV energy domain, it is necessary to cover areas from tens to thousands of square kilometers, what implies the deployment of sparse arrays with distances between antenna stations from tens to thousands of meters.
The hardware and methods for the sparse antenna arrays were developed and successfully tested in AERA and Tunka-Rex~\cite{Abreu:2012pi,Aab:2018ytv,Aab:2015vta,Bezyazeekov:2015ica,Bezyazeekov:2015rpa,Kostunin:2015taa,Bezyazeekov:2018yjw}, which has shown that radio arrays for air-shower detection are ready for the installation on the large areas.
This success have brought a motivation to build large-scale radio arrays with areas of thousands square kilometers.
The first step will be done with the AERA setup which will be extended to the full area of the Pierre Auger Observatory, namely by attaching an antenna station to each water-Cherenkov tank.
The first successful stand-alone array ARIANNA aimed mainly for the neutrino detection is being deployed in Antarctica and has already shown that a self-trigger can be implemented and successfully used in very radio-quiet locations~\cite{Barwick:2016mxm}.
One can see the comparison of the sizes of arrays discussed above in the left panel of Fig.~\ref{fig:all_layouts}.

\section{Towards next-generation radio arrays}
Open problems for the next generation of digital radio arrays are frequency band, self-trigger and size.
Nowadays there are different approaches to the solutions of these problems.
The experiments operating in Antarctica (ARIANNA, ARA~\cite{Allison:2015eky} and ANITA~\cite{Allison:2018cxu}) have lower requirements for a self-trigger system due to lower background level.
Moreover, these experiments use the ice as a target for the ultra-high energy neutrino interaction and measure radio emission from neutrino-induced cascade either inside the ice or from above.
Due to small dimensions of the cascades in the ice, the detection bands for the in-ice detection are shifted towards higher frequencies (from hundred MHz to GHz).
For the air-shower cascades the selection of the band is not obvious, because air-shower pulses feature a broad frequency signature and background contamination varies as a function of frequency.
Thus, the optimal signal-to-noise ratio does not only depend on the width of the band, but also on its lower and upper bounds.
Rough estimation of this band assuming only Galaxy and thermal background were performed~\cite{V.:2017kbm} for a potential surface radio array at IceCube~\cite{Schroder:2018dvb}.
For the regions with higher background (e.g. with anthropogenic RFI), one needs to use more sophisticated methods for a radio self-trigger.
One of them may be the application deep learning algorithms for real-time denoising of radio traces.
An artificial neural network with autoencoder architecture, a most natural selection for a denoising tool based on deep learning, shows promising results on Tunka-Rex background~\cite{Shipilov:2018wph}.
The achievements listed above are planned to be used in the proposed extremely-large scale setup GRAND, a distributed radio array tuned for the detection of very inclined air-showers for neutrinos, cosmic rays and gamma~\cite{Alvarez-Muniz:2018bhp}.
The predicted neutrino detection sensitivities of future experiments are given in the right panel of Fig.~\ref{fig:all_layouts}.

Contrary to ultra-high energy neutrinos, which have unique signature (in-ice and upward-going cascades), the ultra-high energy gammas produce air-shower very similar to ones induced by cosmic rays.
When operating jointly with particle arrays (AERA or radio+IceTop), radio could detect inclined arrays showers induced by gamma in anti-coincidence with particle array, which does not detect such air-shower, since it is absorbed in the atmosphere.
In stand-alone mode gamma/hadron separation can be done by measuring the depth of the shower maximum (high energy gammas penetrate deeper in the atmosphere).

\section{Conclusion}
Nowadays there are two major approaches for detection of the neutrino with radio arrays: detecting ice- and air-showers in radio-quiet regions and building ultra-sparse radio arrays for the detection of inclined air-showers.
The first approach has higher costs per single antenna station (due to logistics and infrastructure), but lower threshold and requirements for the trigger.
Meanwhile the cost efficiency of the second approach has improved (especially with the mass production of antenna stations), but the triggering requires more sophisticated electronics and firmware.
The optimistic estimations predict first detection of EeV neutrinos by future radio arrays in the second half of 2020s.

The detection of the ultra-high energy gammas is possible only with ground arrays and is complicated by the necessity of high-quality gamma-hadron separation.
Nevertheless, already existing radio extensions of the surface arrays (e.g. AERA) increase the sensitivity to gammas arriving from the horizontal direction.
The sensitivity of the future stand-alone arrays to the gamma will mostly depend on their resolution of the depth of shower maximum.

\section*{Acknowledgements} 
The author thanks the organizers of the VLVNT2018 conference for the invitation. 
A part of this work has been supported by the Russian Foundation for Basic Research
(grants No. 16-02-00738, 17-02-00905, 18-32-00460) and by the Helmholtz grant No. HRSF-0027.

\bibliography{references}

\end{document}